\documentstyle[11pt,aaspp4]{article}

\def\mathfont#1{\ifmmode{#1}\else{$#1$}\fi} %for math font  
\def\msun{\ifmmode{\ {\rm M}_\odot}\else{$ {\rm M}_\odot$}\fi}  
\def\msunyr{\ifmmode{\msun \ {\rm yr}^{-1}}\else{$\msun \ {\rm yr}^{-1}$}\fi}  
\def\lae{\mathrel{<\kern-1.0em\lower0.9ex\hbox{$\sim$}}}  
\def\gae{\mathrel{>\kern-1.0em\lower0.9ex\hbox{$\sim$}}}  
\def\ergsec{\mathfont{ {\rm ergs\ s}^{-1}}}

\def\msun{\ifmmode{\ {\rm M}_\odot}\else{$ {\rm M}_\odot$}\fi}  
\def\msunyr{\ifmmode{\msun \ {\rm yr}^{-1}}\else{$\msun \ {\rm yr}^{-1}$}\fi}

\lefthead{McNamara et al.}
\righthead{NEW PERSPECTIVES ON COOLING FLOWS}

\begin{document}

``Constructing the Universe with Clusters of Galaxies,'' IAP, Paris, France, $4-8$ July, 2000. \\

\title{NEW PERSPECTIVES ON COOLING FLOWS AND CLUSTER RADIO SOURCES}

\author{B.R. McNamara\altaffilmark{1,3}, 
M.W. Wise\altaffilmark{2},
L.P. David\altaffilmark{3},
P.E.J. Nulsen\altaffilmark{4},
C.L. Sarazin\altaffilmark{5}}

\altaffiltext{1}{Dept. of Physics \& Astronomy, Ohio University, Athens, OH U.S.A.}
\altaffiltext{2}{MIT/CSR, Cambridge, MA U.S.A.}
\altaffiltext{3}{Harvard-Smithsonian Center for Astrophysics, Cambridge, MA U.S.A.}
\altaffiltext{4}{Engineering Physics, University of Wollongong, Wollongong, AU.}
\altaffiltext{5}{Dept. of Astronomy, University of Virginia, Charlottesville, VA U.S.A.}

\begin{abstract}
We discuss new results from the Chandra X-ray Observatory
and XMM-Newton on cluster cooling flows, emphasizing early
results from our Chandra programs.  We find cooling rates 
reduced by factors of 5--10 compared to those from
earlier missions.  Nevertheless, substantial amounts of keV gas 
appear to be cooling and fueling star formation in central 
dominant cluster galaxies (CDGs).
The structure of the keV thermal gas is remarkably complex,
particularly in regions surrounding the the radio source
and sites of star formation in CDGs.
The radio sources are displacing 
the thermal gas leaving cavities filled with radio emission.
The cavities are apparently supported against the local
gas pressure by
magnetic fields and cosmic rays.  In addition, radio-faint, 
``ghost'' cavities are seen in some clusters.
They may be relics of earlier radio outbursts
rising buoyantly in the intracluster medium.  
The radio sources may reduce the mass deposition rates
by mechanical heating, and by inducing
convective currents that move cool material outward.

\end{abstract}

%\keywords{clusters: globular, peanut --- bosons: bozos}
%\keywords{globular clusters,peanut clusters,bosons,bozos}

\section{The Cooling Flow Problem}

Roughly half of clusters of galaxies have bright cusps of X-ray emission in
their central $\sim 100$ kpc.  The cusps
are associated with so-called cooling flows: regions
of dense gas with short radiative
cooling times (Fabian 1994).  Absent a significant source of heat,
the gas will cool to low temperatures and accrete onto
the cluster's central dominant galaxy (CDG).
The cooling gas will presumably accumulate there in atomic and molecular
clouds and form stars.  Indeed, the likelihood that a CDG
has detectable levels of cold gas and star formation increases
dramatically with the X-ray cooling rates (McNamara 1997; Cardiel et al. 1998). 
However, the cooling rates always exceed 
the observed levels of cold gas and
star formation by large factors.  Cooling rates of hundreds
to as much as two thousand solar masses per year have been 
reported from analyses of $ROSAT$ data.
By comparison, the star formation rates
are generally one to two orders of magnitude less than
the cooling rates.  Furthermore, the systems with the
largest star formation rates appear to have experienced
bursts or episodes of star formation lasting $\lae 100$ Myr or so
(McNamara 1997).  Therefore, unless cooling is likewise episodic,
cold gas would accumulate to the observed levels
in much less than $100$ Myr,  which in turn is much less than the
probable $1-10$ Gyr ages of cooling flows. This troubling
situation leads to the
view that either the matter is accumulating in a dark or
otherwise unusual physical state, cooling flows are young, 
or the cooling rates have been substantially overestimated.

Progress on this problem stagnated through
much of the 1990s because the previous generations
of X-ray telescopes left the cooling regions for the most
part unresolved.  The new Chandra X-ray Observatory  
breathed new life into this problem with 
remarkably crisp spectral images of clusters
resolved on kiloparsec (arcsec)
scales, the same scales on which cold, freshly-accreted gas,
star formation, and powerful radio activity is observed.
In addition, high resolution spectra from Chandra's
High Energy Transmission Grating (HETG) and XMM-Newton's Reflection Grating
Spectrometer (RGS) are probing the spectral lines
responsible for cooling through temperatures
shortward of 1 keV.  In this respect, the
Fe XVII and O VII lines are particularly important (Peterson et al. 2000).

\section{Reduced Cooling Rates from Chandra and XMM-Newton}

It would be premature to comment definitively on the broad trends in cooling
flows, other than to point out that 
the early Chandra and XMM results for several clusters
do not confirm the huge cooling rates derived from 
$ROSAT$ and $Einstein$ data.
The radially decreasing temperature gradients
and short central cooling times $t_{\rm cool}\sim 5\times 10^8$ yr
are certainly observed in several clusters.  However, the spectroscopically
derived cooling rates are factors of 5--10
less than earlier estimates.  For instance, the first cooling flow
observed with Chandra during its orbital verification
phase was the Hydra A cluster.
The {\it spectroscopic} cooling rate derived from spectral imaging on
the ACIS S3, back-illuminated CCD is $\dot M\simeq 35 \msunyr$ 
(McNamara et al. 2000; David et al. 2000), while earlier studies
reported cooling rates of
$\dot M \sim 300-600 \msunyr$ (David et al. 1990; 
Peres et al. 1998, but see Ikebe et al. 1997).  
Similarly, a preliminary analysis of a
Chandra image of Abell 2597 gives $\dot M\simeq 40-80 \msunyr$
(McNamara et al. 2001),
compared to the $ROSAT$ PSPC value of $\dot M\simeq 280-350 \msunyr$ 
(Sarazin \& McNamara 1997; Peres et al. 1998).  Furthermore, 
including a cooling flow component in the spectral models improves
the fits to the data only in the central few tens of kpc, and then
only marginally.  

The only evidence for massive cooling
is found using {morphological} cooling rates, calculated
essentially by dividing the central gas mass
by the cooling time.  This method gives $\dot M$s that are
roughly consistent with results from earlier X-ray missions.
However, the morphological $\dot M$s vastly exceed
the spectroscopic $\dot M$s in the same systems (David et al. 2000).
The inconsistency between these methods
suggests that either the cooling gas is being reheated
or maintained at keV temperatures by some poorly understood 
process (David et al. 2000; Fabian et al. 2000a), 
or the cooling model is wrong. 
  
The most compelling evidence for reduced cooling
is found in  XMM's RGS and Chandra's HETG spectra
for the massive cooling flow in Abell 1835.
Abell 1835 has a putative $\dot M\sim 2300 \msunyr$ cooling flow 
(Allen et al. 1996).  However, high dispersion spectroscopy
with the RGS and HETG provide upper limits of $\dot M \lae 200-300\msunyr$,
based primarily on Fe XVII emission limits 
(Peterson et al. 2000; Wise et al. 2001).  
While this $\dot M$ is
certainly large, the data imply
that either most of the cooling gas is maintained
above $\sim 1$ keV,
or that gas is cooling below 1 keV without an obvious spectroscopic signature  
(Fabian et al 2000a).  

%\subsection{Producing the manuscript}\label{subsec:prod}

\section{Star Formation as the Repository of the Cooling Gas}

Even though the huge cooling rates reported in the last
20 years appear to have subsided, there is evidence for substantial cooling.
Nonetheless, the the mass budgets in the few systems
studied thus far do not balance.
In Hydra A, the strongest spectroscopic evidence for cooling
gas is found in the central 30 or so kpc (David et al. 2000;
McNamara et al. 2000). Within this region sits a
9 kpc diameter circumnuclear disk of young stars and gas (McNamara
1995; Hansen et al. 1995; Melnick et al 1997).  The star formation
history of the disk, i.e. burst or continuous star formation, remains
sketchy.  However, the disk's mass as measured by its
rotation curve and luminosity is
$10^{8\to 9} \msun$. This mass is consistent
with the present rate of cooling throughout the entire volume of
the disk over the Hubble time.  On the other hand, 
if all $35 \msunyr$ of material
cooling within a 70 kpc radius is accreting onto the disk,
it would double its
mass in only 30 Myr. This would be implausible unless
the accretion began recently.  Hydra A's stellar halo is also unusually
blue (Cardiel et al. 1998; McNamara 1995), so distributed star formation
at some level could account for at least some of the cooling gas. 

The CDG in Abell 2597 likewise 
is experiencing  vigorous star formation with a total
mass of $\sim 10^{8\to 9}\msun$ 
(McNamara \& O'Connell 1993; Koekemoer et al. 1999).  
Although we have not
completed our Chandra analysis of Abell 2597 (McNamara et al. 2001), 
the preliminary estimate
of $\dot M\sim 40-80 \msunyr$ would be capable of fueling star formation
for $\sim 10^7$ yr or so depending on the star formation history.
Therefore, while the reduced accretion rates certainly
reduce the mass deficit, most of the cooling mass
remains unseen.

\subsection{X-ray Structure Near the Sites of Star Formation}

The Chandra images of cooling flows we have seen
thus far show surface
brightness irregularities and structure in
the inner few tens of kpc.  Much of the structure
is associated with the radio sources, nebular
emission, and sites of star formation.
For instance, a bright, irregular X-ray structure is associated
with the disk of star formation and nebular emission
in Hydra A (McNamara et al. 2000).  (The structure
surrounding Hydra A's radio lobes is discussed below.)
In addition, bright knots of X-ray emission
(McNamara et al. 2001 in prep.)
accompany the regions of ongoing star formation, nebular emission,
and molecular gas in Abell 2597 (Voit \& Donahue 1997;
Donahue et al. 2000).
These structures are seen in Figure 1 with the radio contours
superposed on the X-ray image.

\begin{figure}
\plotone{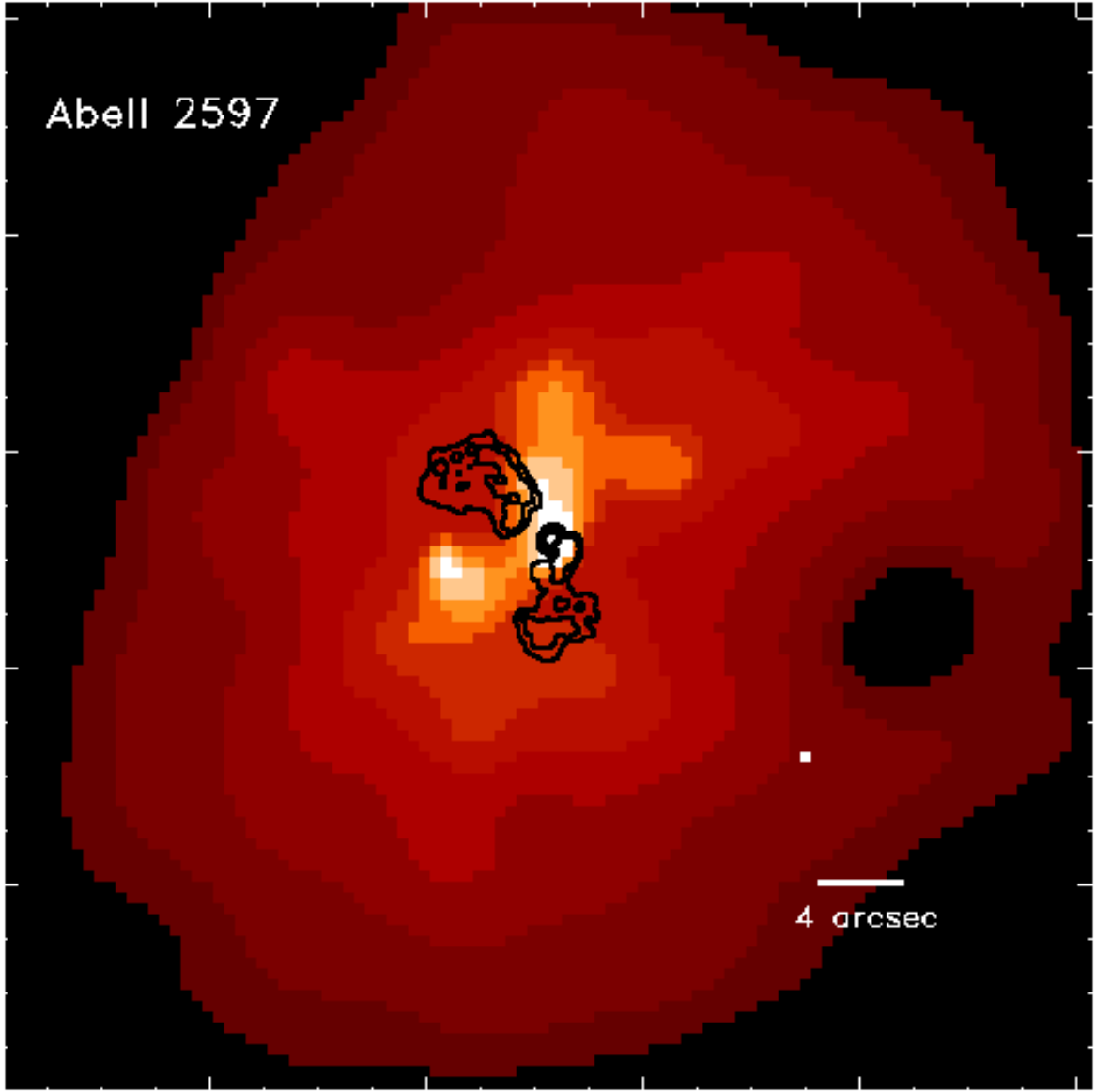}
\caption{40 ksec Chandra image of the center of Abell 2597 (grayscale)
with 8.4 GHz radio contours superposed (McNamara et al. 2001 in prep).
Note the radio-faint hole 17 arcsec to the southwest.  \label{fig1}}
\end{figure}

\placefigure{fig1}

The significance of these spatial correlations are just
beginning to be explored.  However, it has been known
for some time that star formation and nebular emission in cooling flows
are found
preferentially along the edges of the radio sources 
(Heckman et al. 1989; Baum 1992; McNamara 1997; Cardiel et al. 1998).
Star formation occurs in repeated bursts (McNamara \& O'Connell 1993;
Allen 1995; McNamara 1997) and
in many instances, the
radio source seems to be triggering it 
by some mechanism (e.g. De Young 1995; McNamara 1999).
The bright X-ray emission may be from dense, cool
gas that was compressed and displaced by the expanding
radio lobes, or back-flowing material along the
radio lobes.  The fact that the brightest structure is seen in
similar configurations between the radio lobes in several 
objects with a broad range of radio power (e.g. Hydra A,
Abell 2597, M84) supports this general picture.

In Hydra A, the cooling time of the gas reaches a minimum
of $\sim 300-600$ Myr surrounding the blue optical disk
where star formation is observed. Whether this means
star formation is being fueled by cooling is not clear.
A remarkable correlation 
between X-ray structure, star formation, and nebular emission
is seen in the Chandra
image of Abell 1795 (Fabian, this conference).  Abell 1795 has
a tail of $U$-band continuum and H$\alpha$ emission extending 
roughly 70 kpc to the south of the CDG (Cowie et al. 1983;
McNamara et al. 1996).  The Chandra image shows this optical
emission embedded
in a bright X-ray filament.  This feature is unrelated to the radio
source, but may be associated with a group merger or a cooling
filament of gas.

\section{Interactions Between Radio Sources and the ICM}

\subsection{Radio-Bright Cavities in the keV Gas}

Chandra observations show 
a strong interaction between the keV gas
and the powerful radio source Hydra A.  The most striking features
are 30 kpc diameter depressions in the X-ray emission that
coincide with the radio lobes (McNamara et al. 2000).  
Similar surface brightness depressions
in Perseus seen in $ROSAT$ HRI images (B\"ohringer et al 1993) are beautifully
seen in Chandra images (Fabian et al. 2000b).   The radio lobes
appear to be excavating cavities in the keV gas. The 
cavities are presumably supported against the ambient gas pressure
by magnetic fields and cosmic rays.  The density within
the cavities, or bubbles, is much less than their surroundings,
so the bubbles rise
outward by buoyancy (McNamara et al. 2000; Churazov et al. 2000).
Using hydrodynamic models, 
Heinz, Reynolds, \& Begelman (1998) have proposed that 
cavities would form in the ICM during the supersonic expansion
of the radio cocoon.  In this case, the models predict
high-entropy, shocked gas surrounding the radio lobes.
This is not observed. In fact
the gas in the shell around Hydra A's radio lobes is cooler
than the surrounding material.  Fabian et al. (2000b) likewise found cool
gas surrounding NGC 1275's radio cavities.  
David et al. (2000) and Reynolds, Heinz, and Begelman (2000)
have argued that Hydra A is in the weak shock regime of
its expansion, when the
radio lobes are moving at the local sound speed.  

Assuming the radio-bright cavities are in pressure balance with the
surrounding gas, we can derive several interesting properties
(see McNamara et al. 2000; Reynolds, Heinz, and Begelman 2000).  
For instance, the mechanical power required to inflate the
cavities  
$P= pV/t_{\rm b} \simeq 6\times 10^{43}~\ergsec$ is comparable to
Hydra A's total radio power. Here,  
$p$ is the ambient pressure, $V$ is the volume of the cavities,
and $t_{\rm b}$ is the local buoyancy timescale. (A similar
figure is obtained assuming the cavities expanded at the local
sound speed.)  This shows that the radio source is indeed capable
of displacing the keV gas, and that there is little evidence for
radio kinetic luminosity substantially in excess of the luminous
radio power.  Radio-filled cavities are also seen in low radio power
ellipticals such as M84 (Finoguenov \& Jones 2000).  Strong
interactions are evidently occurring in systems covering a wide 
range of gas density and radio power.   

\subsection{Radio-Faint Cavities in the keV Gas}

Cavities in the X-ray emission that lack 
bright radio emission above $\sim 1$ GHz have been discovered in
a few clusters.
For example, Abell 2597 harbors a powerful radio source 
PKS 2322-122 (Sarazin et al. 1995).  The radio source is relatively compact
in the CDG's inner 10 kpc, where it is colliding with 
cold, dusty gas clouds (Sarazin et al. 1995).  
Although the X-ray surface brightness decreases at the
locations of the radio
lobes, cavities are not as obvious as those seen in Hydra A and
Perseus.  However, a prominent, circular cavity is seen 30 kpc
to the southwest of the CDG, and possibly another to the
northeast. The cavity to the southwest is roughly 12 kpc in diameter
and is aligned with the axis of Abell 2597's inner radio jets.
A 15 minute  VLA, A configuration image at 8.44 GHz clearly shows the
inner radio source (Sarazin et al. 1995), but fails to reveal
any emission from the cavities. Although the origin
of these cavities is unknown, they may possibly be the 
outwardly-moving remnants of a previous radio outburst: 
a radio fossil or ``ghost'' (En\ss lin \& Gopal-Krishna 2000;
Medina-Tanco \& En\ss lin 2000).
Assuming they are propelled outward by buoyancy, the causal radio outburst
would have occurred $\sim 100$ Myr ago.  
If the cavities are filled
with magnetic field and cosmic rays with a decaying synchrotron
flux, radio emission may be detectable at lower frequencies. 

Abell 2597 is not unique.  Similar, radio-faint cavities
are seen in Perseus beyond its radio-filled,
inner cavities (Fabian et al. 2000b).  The outer cavities in Perseus
have, however, 
been detected at 74 MHz, which suggests that they are devoid
of energetic electrons (Fabian et al. 2000b).
These bubbles
of magnetic field and cosmic rays are presumably lifted
into the outer regions of clusters where, rejuvenated,
they may contribute
to the formation of cluster-scale radio halos (Kempner \& Sarazin 2000;
En\ss lin \& Gopal-Krishna 2000).

\subsection{Evidence for Repeated Radio Outbursts}

That nearly 70\% of CDGs in cooling flows are radio-bright (Burns 1990)
implies that radio sources live longer than about 1 Gyr,
or they recur with high frequency.  Our interpretation of the cavities
implies the latter, with a recurrence approximately every 100 Myr.
This would have significant implications for
understanding energy feedback to the ICM, the star formation histories
of cooling flows, and the nature and fueling of radio sources.  
Feedback between the central black hole and cooling flow
may be occurring at some level (Tucker \& David 1997; Soker et al. 2000).
Furthermore, star formation occurs in repeated, short
duration bursts (McNamara 1997; Allen 1995; Cardiel et al. 1998),
and there is strong evidence in several objects that the
starbursts were triggered by their radio sources (McNamara 1997).
The starburst ages are generally consistent with the radio-burst
ages implied by the cavities in Hydra A and Abell 2597.
It therefore seems that repeated radio outbursts
affect both the structure and dynamics of the keV gas and
the star formation histories of cooling flows.  However
the physical mechanisms driving these processes
are poorly understood.

%\begin{figure}
%\hspace{1.0in}
%\psfig{figure=xrad.ps,height=4.in} %color
%\psfig{figure=xrad_bw.ps,height=4.in} %black and white
%\caption{40 ksec Chandra image of the center of Abell 2597 (grayscale)
%with 8.4 GHz radio contours superposed (McNamara et al. 2001 in prep).
%Note the radio-faint hole 17 arcsec to the southwest.  
%\label{fig:radish}}
%\end{figure}

\subsection{Radio Heating and Convection in Clusters}

It is too early to tell whether the mechanical energy of radio sources
is heating the ICM and 
lowering the cooling rates (Soker et al. 2000).
On the one hand, Hydra A's entropy profile flattens 
near the center of the cluster,
which suggests the gas is indeed being heated (David et al. 2000).  
Furthermore, the energy input required to substantially reduce the cooling 
rate
is only four times the current radio power.  Therefore,
a plausible amount of mechanical energy above
the radio luminosity would be required to heat the gas.
On the other hand,
the decreasing temperature gradient in the vicinity of the  radio
source is inconsistent with heating, nor is there evidence
for strong shocks (McNamara et al. 2000).
Weak shocks, which are difficult to detect,  may be 
important,  however (David et al. 2000).  So although 
radio sources are obviously having a substantial impact on the dynamics
of the ICM, it is not yet clear that they are heating the 
gas significantly.

Finally, David et al. (2000) 
proposed that convective currents driven by Hydra A's
radio source are removing cooling material from
the core of the cluster and transporting it outward.  
However, the pronounced
metallicity gradient in the cluster's core 
(David et al. 2000; Fukazawa et al. 2000)  is a problem 
for this scenario. Convection should remove a metallicity
gradient, unless the excess metals are rapidly replenished.
Although ongoing star formation is occurring in Hydra A,
the present day star formation rate is too low to replenish 
the metals at the required rate.

\section{Low Luminosity AGN}

Chandra is revealing low luminosity AGN in CDG nuclei.
In addition to being powerful radio sources,
both Hydra A (Sambruna et al. 2000; McNamara et al. 2000)
and Abell 2597 (McNamara et al. 2001) have unresolved nuclear
X-ray sources.  Hydra A's spectrum can be modeled as a heavily absorbed
power law with a total unabsorbed luminosity of
$L_{\rm 2-10 keV}\simeq 1.3\times 10^{42}~\ergsec$, and
an absorbing column of $N_{\rm H}\simeq 4\times 10^{22}~{\rm cm}^2$
(Sambruna et al. 2000).  
Hydra A would have a $\sim 10^9 \msun$ nuclear black hole
were it to lie on the velocity dispersion--black hole mass relation
(Ferrarese \& Merritt 2000; Gebhardt et al. 2000).  The observed nuclear
X-ray flux would then correspond to $\sim 10^{-5}$
times the Eddington luminosity, which is consistent with
its being a low radiative efficiency, advection dominated
accretion flow, i.e. ADAF (Sambruna et al. 2000).  
Furthermore, Hydra A and Abell 2597 both have deep H I absorption
features seen against their nuclear radio sources and jets 
(Taylor 1996; Taylor et al. 1999), which is rare in the general
population of ellipticals (Morganti et al. 2000).  
In Hydra A, the hydrogen column density in front of the
radio core roughly matches the column density derived from the
X-ray spectrum of the AGN.  This implies that the cold material
is within 24 pc of the central black hole (McNamara et al. 2000).  
These properties taken together
support models for AGN fueled by accretion of
cold gas onto a super massive black hole, although the
hot Bondi accretion favored by ADAF models may also be
important (Di Matteo et al. 2000).

%\placetable{tbl-3}
%\placefigure{fig1}

\acknowledgments
BRM thanks Florence Durret and Daniel Gerbal for hosting a wonderful
meeting in Paris, and for their patience with this inexcusably late
contribution.  We thank our Abell 2597 
collaborators, C. O'Dea, S. Baum, M. Donahue, M. Voit, R. O'Connell,
A. Koekemoer, and J. Houck for allowing us to show a picture of Abell 2597 
prior to publication.

%\begin{thebibliography}{}
%\bibitem[Auri\`ere 1982]{aur82} Auri\`ere, M.  1982, \aap,
%    109, 301
%\bibitem[Canizares et al.\ 1978]{can78} Canizares, C. R.,
%    Grindlay, J. E., Hiltner, W. A., Liller, W., and 
%    McClintock, J. E.  1978, \apj, 224, 39

%\section*{References}

\end{document}